\documentstyle[preprint,aps]{revtex}
\tightenlines

\begin{document}
\title{Perturbed Coulombic potentials in Dirac and Klein-Gordon equations}
\author{Omar Mustafa}
\address{Eastern Mediterranean University,}
\date{\today }
\maketitle

\begin{abstract}
A relativistic extension of our pseudo-shifted $\ell $--expansion technique
is presented to solve for the eigenvalues of Dirac and Klein-Gordon
equations. Once more we show the numerical usefulness of its results via
comparison with available numerical integration data.
\end{abstract}

\section{Introduction}

Dirac and Klein-Gordon (KG) equation are not exactly soluble for most of
Lorentz vector ( coupled as the 0-component of the 4-vector) and/or Lorentz
scalar ( added to the mass term) potentials[1-14]. One, therefore, has to
resort to some approximation schemes [1-6]. Yet, in between non-numerical
and purely numerical non-relativistic (Schr\"{o}dinger) and relativistic (KG
and Dirac) wave equations there exists a broad {\em gray zone} of potentials
tractable via various systematic semi-numerical (or semi-analytical)
power-series expansions.

In numerous methodical predecessors of a subset of papers, Mustafa and
co-workers [7] have sought a possibility in the power-law asymptotic
expansions using some {\em small} parameter to solve for Schr\"{o}dinger
equation. It has been noted that the presence of the central spike $\ell
_{d}\,(\ell _{d}+1)\,/\,r^{2}$ (where $\ell _{d}=\ell +(d-3)/2$ and the
dimensions $d\geq 2$) in the radial Schr\"{o}dinger equation, just copies
the effect of the centrifugal and/or centripetal force and immediately
inspires the use of {\em small shifted inverse angular momentum quantum
number}. Their PSLET ( pseudo-perturbative shifted-$\ell $ expansion
technique) has provided persuasive numerical verifications by immediate
comparison of its results with available {\em brute force} numerical data
[7]. PSLET simply consists of using $1/\bar{l}$ as a perturbative expansion
parameter, where $\bar{l}=\ell -\beta _{o}$, $\ell $ is a quantum number,
and $\beta _{o}$ is a suitable shift introduced to avoid the trivial case $%
\ell =0$ [7].pagebreak

In this paper, we extend PSLET\ recipe to solve for Dirac and KG equations
with Lorentz scalar and/or Lorentz vector radially symmetric potentials (in
section 2). In section 3 we apply this relativistic recipe to some exactly
solvable potentials ( e.g., $V(r)=S(r)=-A/r$, $V(r)=-A_{1}/r$ and $S(r)=0$, $%
V(r)=0$ and $S(r)=-A_{2}/r$, and the Dirac oscillator) and non-exactly
solvable (by PSLET) potentials ( e.g., the pure scalar linear, the {\em %
funnel-shapped}, and the power-law potentials) to study the usefulness of
its numerical results. We conclude in section 4.

\section{PSLET recipe for Dirac and KG equations}

The Dirac equation with the Lorentz scalar ( added to the mass term) and
Lorentz vector ( coupled as the 0-component of the 4-vector potential)
potentials reads (in $\hbar =c=1$ units) 
\begin{equation}
\left\{ \vec{\alpha}.\vec{p}+\beta \,\left[ m+S(r)\right] \right\} \,\Psi (%
\vec{r})=\left\{ E-V(r)\right\} \,\Psi (\vec{r}).
\end{equation}
Which decouples into \newline
\begin{equation}
\xi _{1}(r)\,G(r)+\frac{dF(r)}{dr}-\frac{\kappa }{r}F(r)=0,
\end{equation}
\begin{equation}
\xi _{2}(r)\,F(r)-\frac{dG(r)}{dr}-\frac{\kappa }{r}G(r)=0.
\end{equation}
where $\kappa =-\,(\ell +1)$ for $j=\ell +1/2$, $\kappa =\ell $ for $j=\ell
-1/2$, and 
\[
\xi _{1}(r)=E-V(r)-\left[ m+S(r)\right] , 
\]
\[
\xi _{2}(r)=E+m-y(r);\;\;y(r)=V(r)-S(r). 
\]
$E$ is the relativistic energy, and $G(r)$ and $F(r)$ are the large and
small radial components of the Dirac spinor, respectively. In terms of the
large component $G(r),$ equation (2) reads 
\begin{equation}
\left[ \frac{d^{2}}{dr^{2}}-\frac{\kappa (\kappa +1)}{r^{2}}+\frac{1}{\xi
_{2}(r)}\,(\,y^{^{\prime }}(r)\,[\frac{d}{dr}+\frac{\kappa }{r}]\,)+\xi
_{1}(r)\,\xi _{2}(r)\right] G(r)=0,
\end{equation}
where the prime denotes $d/dr.$ It can be shown that with the ansatz 
\begin{equation}
G(r)=\Phi (r)\,\,exp(-p(r)\,/\,2)~;~~\,~~p^{^{\prime }}(r)=y^{^{\prime
}}(r)\,/\,\xi _{2}(r),
\end{equation}
equation (4) reads 
\begin{equation}
\left\{ -\frac{d^{2}}{dr^{2}}+\frac{\kappa (\kappa +1)}{r^{2}}+U(r)-\xi
_{1}(r)\,\xi _{2}(r)\right\} \Phi (r)=0,
\end{equation}
where 
\begin{equation}
U(r)=\frac{y^{^{\prime \prime }}(r)}{2\,\xi _{2}(r)}-\frac{\kappa }{r}\frac{%
\,y^{^{\prime }}(r)}{\,\xi _{2}(r)}+\frac{3}{4}\,\left( \frac{y^{^{\prime
}}(r)}{\xi _{2}(r)}\right) ^{2}.
\end{equation}
Obviously, equation (6) reduces to KG-equation with $\kappa \,(\kappa
+1)=\ell \,(\ell +1)$, for any $\kappa $, if $U(r)$ is set zero. It is
therefore convenient to introduce a parameter $\lambda =0,\,1$ in $U(r)$ so
that $\lambda =0$ and $\lambda =1$ correspond to KG and Dirac equations
respectively. Also, we shall be interested in the problems where the rest
energy $mc^{2}$ is large compared to the binding energy $E_{bind.}=E-mc^{2}$%
. This would manifest the approximation 
\begin{equation}
\frac{1}{\xi _{2}(r)}=\frac{1}{E_{bind.}+2m-y(r)}\simeq \frac{1}{2m}%
-O(1/m^{2}),
\end{equation}
which in turn implies 
\begin{equation}
U(r)=\frac{\lambda }{4\,m}\left[ y^{^{\prime \prime }}(r)-\frac{2\,\kappa
\,y^{^{\prime }}(r)}{r}+\frac{3\,y^{^{\prime }}(r)^{2}}{4\,m}\right] .
\end{equation}

For Coulombic-like potentials (i.e., a Lorentz-vector $V(r)=-A_{1}/r$ and a
Lorentz-scalar $S(r)=-A_{2}/r$ potentials) one may re-scale the potentials
and use the substitutions 
\begin{equation}
V_{r}(r)=V(r)^{2}-\frac{A_{1}^{2}}{r^{2}},
\end{equation}
\begin{equation}
S_{r}(r)=S(r)^{2}-\frac{A_{2}^{2}}{r^{2}},
\end{equation}
to recast equation (6) as 
\begin{equation}
\left[ -\frac{d^{2}}{dr^{2}}+\frac{[\,\bar{l}^{2}+\bar{l}\,(2\,\beta
_{o}+1)+\beta _{o}\,(\beta _{o}+1)]}{r^{2}}+\Gamma (r)+2\,E\,V(r)\right]
\,\Phi (r)=E^{2}\Phi (r).
\end{equation}
where 
\begin{equation}
\Gamma (r)=-V_{r}(r)+S_{r}(r)+2\,m\,S(r)+m^{2}+U(r),
\end{equation}
\begin{equation}
\bar{l}=\ell ^{\prime }-\beta _{o};\,\,\,\,\ell ^{\prime }=-\frac{1}{2}+%
\sqrt{\left( \ell +1/2\right) ^{2}-A_{1}^{2}+A_{2}^{2}},
\end{equation}
and $\beta _{o}$ is a suitable shift to be determined below. Next, we shift
the origin of the coordinate system $x=\bar{l}^{1/2}(r-r_{o})/r_{o}$, where $%
r_{o}$ is currently an arbitrary point to be determined through the
minimization of the leading energy term below. It is therefore convenient to
expand about $x=0$ (i.e., $r=r_{o}$) and use the following expansions 
\begin{equation}
r^{-2}=\sum_{n=0}^{\infty }\,\frac{a_{n}}{r_{o}^{2}}\,x^{n}\,\bar{l}%
^{-n/2};\,\,a_{n}=\left( -1\right) ^{n}\,(n+1),
\end{equation}
\begin{equation}
\Gamma (x(r))=\frac{\bar{l}^{2}}{Q}\,\sum_{n=0}^{\infty }\,b_{n}\,x^{n}\,%
\bar{l}^{-n/2};\,\,b_{n}=\frac{d^{n}\,\Gamma (r_{o})}{dr_{o}^{n}}\,\frac{%
r_{o}^{n}}{n!}
\end{equation}
\begin{equation}
V(x(r))=\frac{\bar{l}}{\sqrt{Q}}\,\sum_{n=0}^{\infty }\,c_{n}\,x^{n}\,\bar{l}%
^{-n/2};\,\,c_{n}=\frac{d^{n}\,V(r_{o})}{dr_{o}^{n}}\,\frac{r_{o}^{n}}{n!}
\end{equation}
\begin{equation}
E=\frac{1}{\sqrt{Q}}\,\sum_{n=-1}^{\infty }\,E^{n}\,\bar{l}^{-n},
\end{equation}
where $Q$ is set equal to $\,\bar{l}^{2}$ at the end of the calculations.
With the above expressions into (12), one may collect all $x$-dependent
terms of order $\bar{l}$ to imply the leading-order approximation for the
energies 
\begin{equation}
E^{\left( -1\right) }=V(r_{o})\pm \sqrt{V(r_{o})^{2}+\Gamma (r_{o})+\frac{Q}{%
r_{o}^{2}}}.
\end{equation}
Which upon minimization, i.e., $dE^{\left( -1\right) }/dr_{o}=0$ and $%
d^{2}E^{\left( -1\right) }/dr_{o}^{2}>0$, yields 
\begin{equation}
2\,Q=h\left( r_{o}\right) +\sqrt{h\left( r_{o}\right) ^{2}-g\left(
r_{o}\right) }
\end{equation}
where 
\begin{equation}
h(r_{o})=r_{o}^{3}\left[ 2\,V(r_{o})\,V^{^{\prime }}(r_{o})+\Gamma
^{^{\prime }}(r_{o})+r_{o}V^{^{\prime }}(r_{o})^{2}\right] ,
\end{equation}
\newline
\begin{equation}
g(r_{o})=r_{o}^{6}\left[ \Gamma ^{^{\prime
}}(r_{o})^{2}+4V(r_{o})V^{^{\prime }}(r_{o})\Gamma ^{^{\prime
}}(r_{o})-4\Gamma (r_{o})V^{^{\prime }}(r_{o})^{2}\right]
\end{equation}
\newline
and primes denote derivatives with respect to $r_{o}$. This implies that $x%
\bar{l}^{-1}$-coefficients vanish, i.e.; 
\begin{equation}
Q\,a_{1}+r_{o}^{2}\,b_{1}+2\,r_{o}^{2}\,E^{\left( -1\right) }\,c_{1}=0.
\end{equation}
Equation (12) therefore reduces to

\begin{eqnarray}
&&\left[ -\frac{d^{2}}{dx^{2}}+\sum_{n=2}^{\infty }T_{n}\,x^{n}\bar{l}%
^{-\left( n-2\right) /2}+\left( 2\,\beta _{o}+1\right) \sum_{n=0}^{\infty
}a_{n}\,x^{n}\,\bar{l}^{-n/2}\right.  \nonumber \\
&&+\beta _{o}\,\left( \beta _{o}+1\right) \,\sum_{n=0}^{\infty
}a_{n}\,x^{n}\,\bar{l}^{-\left( n+2\right) /2}  \nonumber \\
&&\left. +\frac{2\,r_{o}^{2}}{Q}\,\sum_{n=0}^{\infty
}\sum_{p=0}^{n+1}E^{\left( n-p\right) }\left( c_{2p}\,x^{2p}\,\bar{l}%
^{-n}+c_{2p+1}\,x^{2p+1}\bar{l}^{-\left( n+1/2\right) }\right) \right]
\,\Phi _{k,\ell }(x)  \nonumber \\
&=&\left[ \frac{\,r_{o}^{2}}{Q}\,\,\sum_{n=-1}^{\infty
}\sum_{p=-1}^{n+1}E^{\left( n-p\right) }\,E^{\left( p\right) }\,\bar{l}%
^{-\left( n+1\right) }\right] \,\Phi _{k,\ell }(x)
\end{eqnarray}
\newline
\newline
where 
\begin{equation}
T_{n}=a_{n}+\frac{r_{o}^{2}}{Q}\,b_{n}.
\end{equation}
One may now compare equation (24) with the Schr\"{o}dinger equation for the
one-dimensional anharmonic oscillator 
\begin{equation}
\left[ -\frac{d^{2}}{dy^{2}}+\frac{1}{4}\,\omega ^{2}y^{2}+{\Large %
\varepsilon }_{o}+B(y)\right] \,Y_{k}(y)={\Large \mu }_{k}\,Y_{k}(y)
\end{equation}
where $\varepsilon _{o}$ is constant, $B(y)$ is a perturbation like term and 
\begin{equation}
{\Large \mu }_{k}={\Large \varepsilon }_{o}+(k+1/2)\omega
+\sum_{n=1}^{\infty }{\Large \mu }^{\left( n\right) }\,\bar{l}^{-n},
\end{equation}
with $k=0,1,2,\cdots $ and 
\begin{equation}
\omega =\sqrt{12+\frac{2r_{o}^{4}}{Q}\,\Gamma ^{\prime \prime }(r_{o})+\frac{%
4\,r_{o}^{4}}{Q}\,E^{\left( -1\right) }\,V^{\prime \prime }(r_{o}).}
\end{equation}
In a straightforward manner, one can show that 
\begin{equation}
E^{\left( 0\right) }=\frac{Q}{2r_{o}^{2}\left[ E^{\left( -1\right) }-c_{0}%
\right] }\left[ \left( 2\,\beta _{o}+1\right) +\left( k+1/2\right) \omega %
\right] ,
\end{equation}
and choose $\beta _{o}\,$\ so that $E^{\left( 0\right) }=0$ to obtain 
\begin{equation}
\beta _{o}=-\frac{1}{2}\left[ 1+\left( k+1/2\right) \omega \right] .
\end{equation}
equation (24) then becomes

\begin{equation}
\left[ -\frac{d^{2}}{dx^{2}}+\sum_{n=0}^{\infty }{\Huge (}{\Large \upsilon }%
^{\left( n\right) }(x)\,\bar{l}^{-n/2}+J^{\left( n\right) }(x)\,\bar{l}%
^{-n}+K^{\left( n\right) }(x)\,\bar{l}^{-\left( n+1/2\right) }+{\LARGE %
\epsilon }^{\left( n\right) }\,\bar{l}^{-\left( n+1\right) }{\Huge )}\right]
\,\Phi _{k,\ell }(x)=0,
\end{equation}
reak where 
\begin{equation}
{\Large \upsilon }^{\left( 0\right) }(x)=T_{2}\,x^{2}+\left( 2\,\beta
_{o}+1\right) \,a_{0},
\end{equation}

\bigskip

\begin{equation}
{\Large \upsilon }^{\left( 1\right) }(x)=T_{3}\,x^{3}+\left( 2\,\beta
_{o}+1\right) \,a_{1}x,
\end{equation}

\bigskip

\begin{equation}
{\Large \upsilon }^{\left( n\right) }(x)=T_{n+2}\,x^{n+2}+\left( 2\,\beta
_{o}+1\right) \,a_{n}\,x^{n}+\beta _{o}\,(\beta
_{o}+1)\,a_{n-2}\,x^{n-2};\,\ n\geq 2,
\end{equation}

\bigskip

\begin{equation}
J^{\left( n\right) }(x)=\left( \frac{2\,r_{o}^{2}}{Q}\right)
\sum_{p=0}^{n+1}E^{\left( n-p\right) }\,c_{2p}\,x^{2p},
\end{equation}

\bigskip

\begin{equation}
K^{\left( n\right) }(x)=\left( \frac{2\,r_{o}^{2}}{Q}\right)
\sum_{p=0}^{n+1}E^{\left( n-p\right) }\,c_{2p+1}\,x^{2p+1}\,
\end{equation}

\bigskip

\begin{equation}
{\Huge \epsilon }^{\left( n\right) }=\left( \frac{\,r_{o}^{2}}{Q}\right)
\sum_{p=-1}^{n+1}E^{\left( n-p\right) }\,E^{\left( p\right) }.
\end{equation}
Now we may closely follow PSLET recipe for the $k$-nodal wavefunction and
define

\begin{equation}
\Phi _{k,\ell }(x)=F_{k,\,\ell }(x)\,\exp (U_{k,\,\ell }(x))
\end{equation}
where 
\begin{equation}
F_{k,\,\ell }(x)=x^{k}+\sum_{n=0}^{\infty }\sum_{p=0}^{k-1}A_{p,k}^{\left(
n\right) }\,x^{p}\,\bar{l}^{-n/2},
\end{equation}
\begin{equation}
U_{k,\,\ell }^{\prime }(x)=\sum_{n=0}^{\infty }\left( U_{k,\,\ell }^{\left(
n\right) }(x)\,\,\bar{l}^{-n/2}+G_{k,\,\ell }^{\left( n\right) }(x)\,\,\bar{l%
}^{-\left( n+1\right) /2}\right) ,
\end{equation}
with 
\begin{equation}
U_{k,\,\ell }^{\left( n\right)
}(x)=\sum_{p=0}^{n+1}D_{p,n,k}\,x^{2p-1};\;\;D_{0,n,k}=0,
\end{equation}

\bigskip

\begin{equation}
G_{k,\,\ell }^{\left( n\right) }(x)=\sum_{p=0}^{n+1}C_{p,n,k}\,x^{2p}.
\end{equation}
Equation (31) then reads\bigskip

\begin{eqnarray}
&&F_{k,\,\ell }(x)\sum_{n=0}^{\infty }{\huge [}{\Large \upsilon }^{\left(
n\right) }(x)\,\bar{l}^{-n/2}+J^{\left( n\right) }(x)\,\bar{l}%
^{-n}+K^{\left( n\right) }(x)\,\bar{l}^{-\left( n+1/2\right) }-{\Huge %
\epsilon }^{\left( n\right) }\,\bar{l}^{-\left( n+1\right) }{\huge ]} 
\nonumber \\
&&-F_{k,\,\ell }(x)\,{\huge [}U_{k,\,\ell }^{^{\prime \prime
}}(x)+U_{k,\,\ell }^{^{\prime }}(x)U_{k,\,\ell }^{^{\prime }}(x){\huge ]}%
-2\,F_{k,\,\ell }^{\prime }(x)\,U_{k,\,\ell }^{^{\prime }}(x)-F_{k,\,\ell
}^{^{\prime \prime }}(x)=0
\end{eqnarray}
where primes denote derivatives with respect to $x$. One may also eliminate $%
\,\bar{l}$-dependance from equation (43) to obtain four exactly solvable
recursive relations ( see Appendix for details). Once $r_{o}$ is determined,
through equation (20), one may then calculate the energy eigenvalues and
eigenfunctions from the knowledge of $C_{p,n,k}$, $D_{p,n,k}$ and $%
A_{p,k}^{\left( n\right) }$ in a hierarchical manner.

\section{Illustrative examples}

In this section we illustrate the applicability of the above relativistic
PSLET recipe through some examples covering Dirac and KG equations.

\subsection{An equally mixed Coulomb potentials}

For an equally mixed Coulombic potentials, i.e. $V(r)=S(r)=-A/r$, $U(r)$ in
(9) vanishes. Consequently $\Gamma (r)=-2mA/r+m^{2}$, $Q=-A^{2}+2mAr$, $%
\omega =2$, $\beta _{o}=-\left( k+1\right) $, $r_{o}=\left[ \left( \ell
^{\prime }+k+1\right) ^{2}+A^{2}\right] \,/\left( 2mA\right) $, and the
leading-order approximation reads 
\[
E^{\left( -1\right) }=m\,\left[ 1-\frac{2\,A^{2}}{\left( k+\ell +1\right)
^{2}+A^{2}}\right] , 
\]
which is the well known exact result for the generalized Dirac- and
KG-Coulomb problems, where higher-order corrections vanish identically.

\subsection{Vector Coulomb or scalar Coulomb potential}

For $V(r)=-A_{1}/r$ and $S(r)=0$ or $V(r)=0$ and $S(r)=-A_{2}/r$ in KG
equation one would obtain the well known exact results 
\[
E^{\left( -1\right) }=m\left[ 1+\frac{A_{1}^{2}}{n_{1}^{2}}\right]
^{-1/2};\;\;n_{1}=k+\frac{1}{2}+\sqrt{\left( \ell +1/2\right) ^{2}-A_{1}^{2}}
\]
or 
\[
E^{\left( -1\right) }=\pm \,m\left[ 1-\frac{A_{2}^{2}}{n_{2}^{2}}\right]
^{1/2};\;\;n_{2}=k+\frac{1}{2}+\sqrt{\left( \ell +1/2\right) ^{2}+A_{2}^{2}} 
\]
respectively. Again higher-order corrections vanish identically.

\subsection{Dirac oscillator}

Following the work of Romero et al [13], the Dirac oscillator [14]
eigenvalue problem ( see equation (30) in [13]) reduces to 
\[
\left[ -\frac{d^{2}}{dr^{2}}+\frac{\Lambda (\Lambda +\epsilon \,\beta )}{%
r^{2}}+m^{2}B^{2}r^{2}+m^{2}+mB\left( \epsilon \left[ 2j+1\right] -\beta
\right) \right] \,\Phi (r)=E^{2}\Phi (r), 
\]
where $\Lambda =j+1/2$, $B$ is the oscillator frequency, and $\epsilon =\pm
1 $. In this case our $\Gamma (r)=m^{2}B^{2}r^{2}+m^{2}+mB\left( \epsilon %
\left[ 2j+1\right] -\beta \right) $, $\omega =4$, $r_{o}^{2}=$ $\bar{l}\,/mB$%
, and our leading term reads 
\[
E^{\left( -1\right) }=\pm \,\left[ 2mB\left( 2k+\ell +3/2\right)
+m^{2}+mB\left( \epsilon \left[ 2j+1\right] -\beta \right) \right] ^{1/2} 
\]
with heigher-order terms identical zeros. Thus, if we take $N=2k+\ell $ (
the harmonic oscillator principle quantum number) we come out with the exact
Dirac oscillator's closed form solution (see equation (35) in [13]) 
\[
E^{2}-m^{2}=\,\left[ 2mB\left( N+3/2\right) +mB\left( \epsilon \left[ 2j+1%
\right] -\beta \right) \right] . 
\]
\bigskip

\subsection{Pure scalar linear potential}

A pure scalar linear potential, i.e., $S(r)=Ar$ and $V(r)=0$, is precisely a
quark confining potential. It has been used by Gunion and Li [5] in Dirac
equation to find, numerically, part of Dirac $J/\Psi $ mass spectra.

Obviously, for this potential equation (20) has to be solved numerically.
Then one can proceed to obtain the mass spectra for $A=0.137\,GeV^{2}$, $%
m=1.12\,GeV$, $\kappa =-\left( \ell +1\right) $, and $\kappa =\ell $ through
the prescription $M=2E$.

In tables 1 and 2 we report our results for $J/\Psi $ mass ( in $GeV$) for $%
\kappa =-\left( \ell +1\right) $ and $\kappa =\ell $, respectively. To show
the trends of convergence of our results, we report them as $M(N)=2E(N)$,
with $N$ denoting the number of corrections added to the leading-order
approximation $E^{\left( -1\right) }$. Our results are also compared with
the numerically predicted ones of Gunion and Li [5]. Evidently, the accuracy
and trend of convergence are satisfactory.

\subsection{{\em Funnel-shaped} potential}

The {\em funnel-shaped }potential is widely used in quarkonium physics. It
has both vector and components, $V(r)$ and $S(r)$, respectively.

In Dirac equation, Stepanov and Tutik [4] have used numerical integrations
and $\hbar $-expansion formalism without the traditional conversion of Dirac
equation into a Schr\H{o}dinger-like form (unlike what we have already done
in section II above). They have obtained the Charmonium masses for $%
V(r)=-2\alpha /3r$ and $S(r)=br/2$, where $m=1.358\,GeV$, $\alpha =0.39$,
and $b=0.21055\,GeV^{2}$.

In table 3 we show our results for the Charmonium masses for $\kappa =\ell $
and compare them with those of numerical integration and $\hbar $-expansions
of Stepanov and Tutik [4]. They are in good agreement and the trend of
convergence of our results is also satisfactory. However, in table 4 we
report the Charmonium masses, for $\kappa =-\left( \ell +1\right) $.
Therein, we only list our results where the mass series and Pad\'{e}
approximants stabilize.

In KG equation Kobylinsky, Stepanov, and Tutik [3] have used $V(r)=-a/r$ and 
$S(r)=b\,r$ with $m=1.370\,GeV$, $b=0.10429\,GeV^{2}$ and $a=0.26\,$\ to
obtain the energies for this {\em funnel-shaped }potential. They have also
used $\hbar $-expansions and numerical integrations. In table 5 we list our
results and compare them with those of $\hbar $-expansions, $E_{\hbar }$,
and numerical integrations, $E_{num}$, reported in [3]. They are in
excellent agreement.

\subsection{Power-law potential}

An equally mixed scalar and vector power-law potential 
\[
V(r)=S(r)=Ar^{\nu }+B_{o} 
\]
where $\nu =0.1$ ( the Martin potential [6] ) and $A>0$ is found most
successful in describing the entire light and heavy meson spectra in the
Dirac equation (cf. e.g., Martin 1980, 1981, and Jena and Tripati in [6]).
Once such potential setting is used in Eq.(6) along with the substitution $%
q=r/\varrho $, with 
\[
\varrho =\left[ 2\left( E+m\right) A\right] ^{-1/\left( \nu +2\right) }, 
\]
one gets a simple Schr\"{o}dinger-type form 
\begin{equation}
\left[ -\frac{d^{2}}{dq^{2}}+\frac{\ell \left( \ell +1\right) }{q^{2}}%
+q^{\nu }\right] \,\Omega \left( q\right) =\check{E}\,\Omega \left( q\right)
,
\end{equation}
where 
\begin{equation}
\check{E}=\left( E-m-2B_{o}\right) \left[ \left( E+m\right) \left( 2A\right)
^{-2/\nu }\right] ^{\nu /\left( \nu +2\right) }.
\end{equation}
Therefore, one better solve (44) for $\check{E}\,(N)$ and then to find the
Dirac quark binding energies $E$ from (45). In table 6 we compare PSLET
results with those obtained numerically, $E_{num},$ by Jena and Tripati [6].
The results from the shifted $1/N$ - expansion technique by Roy and
Roychoudhury [6], $E_{1/N},$ are also listed.

\section{Concluding remarks}

In this work we presented a straightforward extension of our PSLET recipe
[7] to solve for the eigenvalues of Dirac and KG equations, with Lorentz
vector and/or Lorentz scalar potentials. We have, again, documented (through
tables 1-5) the usefulness of this recipe by immediate comparisons with
available numerical integration data.

Nevertheless, one should notice that our results from KG equation are only
partially better , compared with those from $\hbar $-expansions and
numerical integrations, than our results from Dirac equation. The reason is
obviously, and by large, manifested by our approximation in equation (8).
For Klein-Gordon equation $\lambda =0$ in (9) while for Dirac equation $%
\lambda =1$.

In the process, moreover, there still remain some issues of delicate nature.
Namely, one can not obtain (using our PSLET above) the exact eigenvalues for 
$V(r)=-A_{1}/r$ and $S(r)=0$, $V(r)=0$ and $S(r)=-A_{2}/r$, or even for $%
V(r)=-A_{1}/r$ and $S(r)=-A_{2}/r$ in Dirac equation.

The remedy seems to be feasible in a sort of combination between the current
relativistic PSLET and a similarity transformation (cf., e.g. ref.[8] and
related references therein). Preliminary results show that if $%
S(r)\longrightarrow -A_{2}/r+S_{o}(r)$ and $V(r)\longrightarrow
-A_{1}/r+V_{o}(r)$ in (1) such that $S_{o}(r)\longrightarrow 0$ and $%
V_{o}(r)\longrightarrow 0$ as $r\longrightarrow 0$, then a similarity
transformation could accompany our relativistic PSLET to obtain exact
results for the generalized Dirac-Coulombic problem and better results for
potentials of the type $S(r)\longrightarrow -A_{2}/r+S_{o}(r)$ and $%
V(r)\longrightarrow -A_{1}/r+V_{o}(r)$. That is, one may carefully follow
Mustafa's work [8] to obtain 
\begin{equation}
\left[ -\frac{d^{2}}{dr^{2}}+\frac{(\,\gamma ^{2}+s\,\gamma )}{r^{2}}%
+U(r)+m^{2}+\frac{2}{r}\,\left( A_{2}\,m(r)+E(r)\,A_{1}\right) \right]
\,\Phi (r)=E^{2}\Phi (r).
\end{equation}
where $\gamma =\sqrt{\kappa ^{2}-A_{1}^{2}+A_{2}^{2}}$ , $s=\pm 1$, $%
E(r)=E-V_{o}(r)$, $m(r)=m+S_{o}(r)$, and $U(r)\longrightarrow 0$ as $%
V_{o}(r)\longrightarrow 0$ and $S_{o}(r)\longrightarrow 0$ (hence, $%
m(r)\longrightarrow m$ and $E(r)\longrightarrow E$). Therefore, one would
replace our $\ell ^{\prime }$, in (12), by $\ell ^{\symbol{126}}=-1/2+\gamma
+s\,/\,2$ and obtain ( following PSLET\ recipe above) for (48), $\Gamma
(r_{o})=m^{2}-2mA_{2}r_{o}$, $V(r)=-A_{1}/r$, $\omega =2$, $\beta
_{o}=-\left( k+1\right) $, and $\bar{l}=\ell ^{\symbol{126}}-\beta
_{o}=(k+1/2+s\,/\,2+\,\gamma )$. In turn, equation (23) yields 
\begin{equation}
4Q\left[ Q-2mA_{2}r_{o}+A_{1}^{2}\right] +4m^{2}r_{o}^{2}\left[
A_{2}^{2}-A_{1}^{2}\right] =0
\end{equation}
to solve for $r_{o}$. This would lead to 
\begin{equation}
E^{\left( -1\right) }=m\left[ 1+\frac{A_{1}^{2}}{Q}\right]
^{-1/2};\;\;Q=\left( k+1/2+s\,/\,2+\,\sqrt{\kappa ^{2}-A_{1}^{2}}\right) ^{2}
\end{equation}
and 
\begin{equation}
E^{\left( -1\right) }=\pm \,m\left[ 1-\frac{A_{2}^{2}}{Q}\right]
^{1/2};\;\;Q=\left( k+1/2+s\,/\,2+\,\sqrt{\kappa ^{2}+A_{2}^{2}}\right) ^{2}
\end{equation}
for $A_{2}=0$, $A_{1}\neq 0$ and $A_{1}=0$, $A_{2}\neq 0$, respectively. It
should be noted that these are the well known exact results (cf., e.g.; ref
[8]) with the heigher-order terms vanish identically.

Before we conclude it should be noted that if our results are to be
generalized to $d$-dimensions we may incorporate interdimensional
degeneracies associated with the isomorphism between the angular momentum $%
\ell $ and dimensionality $d$ (cf., e.g., Mustafa and Odeh (2000) [7]). This
would replace our $\kappa $ by $\kappa _{d}=s(2j+d-2)/2$, where $\ell
_{d}=\ell +(d-3)/2$. In this way we reproduce Stepanov and Tutik's [4] and
Dong's [9] results in $d$-dimensions.

Finally, the above has been a very limited review and a number of other
useful and novel approaches such as those of Franklin [10, and references
therein], Njock et al [11, and references therein], $\cdots $etc., have not
been touched on.

\newpage \bigskip {\Large Appendix A: Further algebraic simplifications for
PSLET\ relativistic recipe}

Eliminating $\bar{l}$-dependence, equation (42) can be simplified into four
recursive relations to read 
\begin{equation}
k(k-1)\,x^{k-2}+T_{k,\ell }^{\left( 0\right) }(x)-N_{k,\ell }^{\left(
0\right) }(x)=0
\end{equation}
\begin{equation}
T_{k,\ell }^{\left( 1\right) }(x)+S_{k,\ell }^{\left( 0\right)
}(x)-O_{k,\ell }^{\left( 0\right) }(x)=0
\end{equation}
and for $n\geq 0$%
\begin{equation}
T_{k,\ell }^{\left( 2n+2\right) }(x)-N_{k,\ell }^{\left( n+1\right)
}(x)+S_{k,\ell }^{\left( 2n+1\right) }(x)+M_{k,\ell }^{\left( 2n\right)
}(x)+\Lambda _{k,\ell }^{\left( n\right) }(x)=0
\end{equation}
\begin{equation}
T_{k,\ell }^{\left( 2n+3\right) }(x)+S_{k,\ell }^{\left( 2n+2\right)
}(x)-O_{k,\ell }^{\left( n+1\right) }(x)+M_{k,\ell }^{\left( 2n+1\right)
}(x)+{\Large \zeta }_{k,\ell }^{\left( n\right) }(x)=0
\end{equation}
where 
\begin{eqnarray}
T_{k,\ell }^{\left( n\right) }(x) &=&L_{k,\ell }^{\left( n\right) ^{\prime
\prime }}(x)+2\,k\,x^{k-1}U_{k,\,\ell }^{\left( n\right) }(x)  \nonumber \\
&&+\sum_{p=0}^{n}2L_{k,\ell }^{\left( p\right) ^{\prime }}(x)\,U_{k,\,\ell
}^{\left( n-p\right) }(x)+x^{k}\,\left[ U_{k,\,\ell }^{\left( n\right)
^{\prime }}(x)+R_{k,\,\ell }^{\left( n\right) }(x)-{\Large \upsilon }%
^{\left( n\right) }(x)\right]  \nonumber \\
&&+\sum_{p=0}^{n}L_{k,\ell }^{\left( p\right) }(x)\left( U_{k,\,\ell
}^{\left( n-p\right) ^{\prime }}(x)+R_{k,\,\ell }^{\left( n-p\right) }(x)-%
{\Large \upsilon }^{\left( n-p\right) }(x)\right)
\end{eqnarray}
\begin{eqnarray}
S_{k,\ell }^{\left( n\right) }(x) &=&2\,k\,x^{k-1}G_{k,\,\ell }^{\left(
n\right) }(x)+\sum_{p=0}^{n}2L_{k,\ell }^{\left( p\right) ^{\prime
}}(x)\,G_{k,\,\ell }^{\left( n-p\right) }(x)  \nonumber \\
&&+x^{k}\,\left[ G_{k,\,\ell }^{\left( n\right) ^{\prime }}(x)+Q_{k,\,\ell
}^{\left( n\right) }(x)\right] +\sum_{p=0}^{n}L_{k,\ell }^{\left( p\right)
}(x)\left( G_{k,\,\ell }^{\left( n-p\right) ^{\prime }}(x)+Q_{k,\,\ell
}^{\left( n-p\right) }(x)\right)
\end{eqnarray}
\begin{equation}
M_{k,\ell }^{\left( n\right) }(x)=x^{k}\,P_{k,\ell }^{\left( n\right)
}\left( x\right) +\sum_{p=0}^{n}L_{k,\ell }^{\left( p\right)
}(x)\,\,P_{k,\ell }^{\left( n-p\right) }\left( x\right)
\end{equation}
\begin{equation}
N_{k,\ell }^{\left( n\right) }(x)=x^{k}\,J^{\left( n\right) }\left( x\right)
+\sum_{p=0}^{n}L_{k,\ell }^{\left( p\right) }(x)\,\,J^{\left( n-p\right)
}\left( x\right)
\end{equation}
\begin{equation}
O_{k,\ell }^{\left( n\right) }(x)=x^{k}\,K^{\left( n\right) }\left( x\right)
+\sum_{p=0}^{n}L_{k,\ell }^{\left( p\right) }(x)\,\,K^{\left( n-p\right)
}\left( x\right)
\end{equation}
\begin{equation}
\Lambda _{k,\ell }^{\left( n\right) }(x)=x^{k}\,{\Huge \epsilon }^{\left(
n\right) }+\sum_{p=0}^{n}L_{k,\ell }^{\left( 2p\right) }(x)\,{\Huge \epsilon 
}^{\left( n-p\right) }
\end{equation}
\begin{equation}
{\Large \zeta }_{k,\ell }^{\left( n\right) }(x)=\sum_{p=0}^{n}L_{k,\ell
}^{\left( 2p+1\right) }(x)\,{\Huge \epsilon }^{\left( n-p\right) }
\end{equation}
\begin{equation}
R_{k,\ell }^{\left( n\right) }(x)=\sum_{p=0}^{n}U_{k,\,\ell }^{\left(
p\right) }(x)U_{k,\,\ell }^{\left( n-p\right) }(x)
\end{equation}
\begin{equation}
P_{k,\ell }^{\left( n\right) }(x)=\sum_{p=0}^{n}G_{k,\,\ell }^{\left(
p\right) }(x)G_{k,\,\ell }^{\left( n-p\right) }(x)
\end{equation}
\begin{equation}
Q_{k,\ell }^{\left( n\right) }(x)=\sum_{p=0}^{n}2U_{k,\,\ell }^{\left(
p\right) }(x)G_{k,\,\ell }^{\left( n-p\right) }(x)
\end{equation}
\begin{equation}
L_{k,\ell }^{\left( n\right) }(x)=\sum_{p=0}^{k-1}A_{k,\,p}^{\left( n\right)
}(x)\,x^{p}
\end{equation}

\begin{description}
\item  {\bf Table1: }PSLET results for part of Dirac $J/\Psi $ spectra (in
GeV) for $S(r)=Ar,$ $A=0137GeV^{2},$ $V(r)=0,$ $\kappa =-(\ell +1),$ and $%
m=1.12GeV.$ Where $M(N)=2E(N)$ with $N$ denoting the number of corrections
added to the leading - order term $E^{(-1)}$ and $M_{num}$ are the numerical
values reported by Gunion and Li [5].

\begin{tabular}{cccccc}
\hline\hline
$k$ & $\ M(N)$ & $\ell =0$ & $\ \ell =1$ & $\ \ell =2$ & $\ \ell =3$ \\ 
\hline\hline
$
\begin{tabular}{c}
$0$%
\end{tabular}
$ & $
\begin{tabular}{c}
$M(1)$ \\ 
$M(2)$ \\ 
$M(3)$ \\ 
$M(4)$ \\ 
$M(5)$ \\ 
${\bf \vdots }$ \\ 
$M(14)$ \\ 
$M_{num}$%
\end{tabular}
$ & 
\begin{tabular}{c}
3.0919 \\ 
3.0961 \\ 
3.0963 \\ 
3.0961 \\ 
3.0961 \\ 
${\bf \vdots }$ \\ 
3.0961 \\ 
3.103
\end{tabular}
& 
\begin{tabular}{c}
3.43078 \\ 
3.43252 \\ 
3.43259 \\ 
4.43256 \\ 
3.43256 \\ 
${\bf \vdots }$ \\ 
3.43256 \\ 
3.442
\end{tabular}
& 
\begin{tabular}{c}
3.711960 \\ 
3.712914 \\ 
3.712947 \\ 
3.712940 \\ 
3.712939 \\ 
$\vdots $ \\ 
3.712939 \\ 
3.725
\end{tabular}
& 
\begin{tabular}{c}
3.9581219 \\ 
3.9587277 \\ 
3.9587465 \\ 
3.9587436 \\ 
3.9587434 \\ 
${\bf \vdots }$ \\ 
3.9587434 \\ 
3.973
\end{tabular}
\\ \hline\hline
\begin{tabular}{c}
$2$%
\end{tabular}
& 
\begin{tabular}{c}
$M(1)$ \\ 
$M(2)$ \\ 
$M(3)$ \\ 
$M(4)$ \\ 
$M(5)$ \\ 
$M(6)$ \\ 
$M(7)$ \\ 
$\ {\bf \vdots }$ \\ 
$M(14)$ \\ 
$M_{num}$%
\end{tabular}
& 
\begin{tabular}{l}
4.131 \\ 
4.142 \\ 
4.148 \\ 
4.150 \\ 
4.151 \\ 
4.152 \\ 
4.152 \\ 
$\ \ {\bf \vdots }$ \\ 
4.152 \\ 
4.158
\end{tabular}
& 
\begin{tabular}{l}
\ 4.3395 \  \\ 
\ 4.3468 \\ 
\ 4.3502 \\ 
\ 4.3515 \\ 
\ 4.3519 \\ 
\ 4.3520 \\ 
\ 4.3521 \\ 
$\ \ \ {\bf \vdots }$ \\ 
\ 4.3521 \\ 
\ 4.36
\end{tabular}
& 
\begin{tabular}{l}
4.5325 \\ 
4.5378 \\ 
4.5401 \\ 
4.5408 \\ 
4.5410 \\ 
4.5411 \\ 
4.5411 \\ 
$\ \ {\bf \vdots }$ \\ 
4.5411 \\ 
4.551
\end{tabular}
& 
\begin{tabular}{l}
4.71334 \\ 
4.71739 \\ 
4.71905 \\ 
4.71950 \\ 
4.71961 \\ 
4.71965 \\ 
4.71966 \\ 
$\ \ \ {\bf \vdots }$ \\ 
4.71966 \\ 
4.732
\end{tabular}
\\ \hline\hline
\end{tabular}
\newline
\newline
\thinspace \newpage

\item  {\bf Table 2: }Same as table 1 for $\kappa =\ell .$\newline

\item  
\begin{tabular}{cccccc}
\hline\hline
$k$ & $M(N)$ & $\ell =1$ & $\ell =2$ & $\ell =3$ & $\ell =4$ \\ \hline\hline
\begin{tabular}{c}
$0$%
\end{tabular}
\  & 
\begin{tabular}{c}
$M(1)$ \\ 
$M(2)$ \\ 
$M(3)$ \\ 
$M(4)$ \\ 
$M(5)$ \\ 
${\bf \vdots }$ \\ 
$M(14)$ \\ 
$M_{num}$%
\end{tabular}
& 
\begin{tabular}{c}
3.47090 \\ 
3.47183 \\ 
3.47188 \\ 
3.47186 \\ 
3.47186 \\ 
${\bf \vdots }$ \\ 
3.47186 \\ 
3.47
\end{tabular}
& 
\begin{tabular}{c}
3.760125 \\ 
3.760677 \\ 
3.760700 \\ 
3.760696 \\ 
3.760695 \\ 
${\bf \vdots }$ \\ 
3.760695 \\ 
3.757
\end{tabular}
& $
\begin{tabular}{c}
4.0111817 \\ 
4.0115488 \\ 
4.0115615 \\ 
4.0115597 \\ 
4.0115595 \\ 
${\bf \vdots }$ \\ 
4.0115595 \\ 
4.006
\end{tabular}
$ & 
\begin{tabular}{c}
4.2364739 \\ 
4.2367365 \\ 
4.2367444 \\ 
4.2367435 \\ 
4.2367435 \\ 
${\bf \vdots }$ \\ 
4.2367435 \\ 
4.23
\end{tabular}
\\ \hline\hline
\begin{tabular}{c}
1
\end{tabular}
& 
\begin{tabular}{c}
$M(1)$ \\ 
$M(2)$ \\ 
$M(3)$ \\ 
$M(4)$ \\ 
$M(5)$ \\ 
$M(6)$ \\ 
${\bf \vdots }$ \\ 
$M(14)$ \\ 
$M_{num}$%
\end{tabular}
& 
\begin{tabular}{c}
3.9570 \\ 
3.9624 \\ 
3.9640 \\ 
3.9644 \\ 
3.9646 \\ 
3.9646 \\ 
${\bf \vdots }$ \\ 
3.9646 \\ 
3.965
\end{tabular}
& 
\begin{tabular}{c}
4.19083 \\ 
4.19451 \\ 
4.19537 \\ 
4.19557 \\ 
4.19561 \\ 
4.19563 \\ 
${\bf \vdots }$ \\ 
4.19563 \\ 
4.194
\end{tabular}
& 
\begin{tabular}{c}
4.40310 \\ 
4.40578 \\ 
4.40631 \\ 
4.40642 \\ 
4.40644 \\ 
4.40644 \\ 
${\bf \vdots }$ \\ 
4.40644 \\ 
4.403
\end{tabular}
& 
\begin{tabular}{c}
4.599080 \\ 
4.601126 \\ 
4.601482 \\ 
4.601542 \\ 
4.601552 \\ 
4.601554 \\ 
${\bf \vdots }$ \\ 
4.601554 \\ 
4.597
\end{tabular}
\\ \hline\hline
\end{tabular}

\item  \newpage

\item  {\bf Table 3: }PSLET \ Charmonium masses $M(N)=2E(N)$ for the {\em %
funnel-shaped} potential, $V(r)=-2\alpha /3r$ and $S(r)=br/2$, with $%
m=1.358\,GeV$, $b=0.21055\,GeV^{2}$, $\alpha =0.39$ and $\kappa =\ell .$ The
quantum numbers in parentheses are ( $k,\ell $). $M_{num}$ is the numerical
integration and $M_{\hbar }$ is the $\hbar $-expansion result ( up to the
tenth-order correction) reported by Stepanov and Tutik [4].

\item  
\begin{tabular}{ccccccc}
\hline\hline
$M(N)$ & $(0,1)$ & $(0,2)$ & $(1,1)$ & $(1,3)$ & $(2,1)$ & $(2,3)$ \\ 
\hline\hline
\begin{tabular}{c}
$M(1)$ \\ 
$M(2)$ \\ 
$M(3)$ \\ 
$M(4)$ \\ 
$M(5)$ \\ 
$M(6)$ \\ 
$M(7)$ \\ 
$M(8)$ \\ 
${\bf \vdots }$ \\ 
$M(14)$ \\ 
$M_{\hbar }$ \\ 
$M_{num}$%
\end{tabular}
& 
\begin{tabular}{c}
3.5071 \\ 
3.5062 \\ 
3.5056 \\ 
3.5055 \\ 
3.5055 \\ 
3.5055 \\ 
3.5055 \\ 
3.5055 \\ 
${\bf \vdots }$ \\ 
3.5055 \\ 
3.4998 \\ 
3.4998
\end{tabular}
& 
\begin{tabular}{c}
3.8012 \\ 
3.8007 \\ 
3.8006 \\ 
3.8005 \\ 
3.8005 \\ 
3.8005 \\ 
3.8005 \\ 
3.8005 \\ 
${\bf \vdots }$ \\ 
3.8005 \\ 
3.7974 \\ 
3.7974
\end{tabular}
& 
\begin{tabular}{c}
3.966 \\ 
3.963 \\ 
3.961 \\ 
3.959 \\ 
3.959 \\ 
3.958 \\ 
3.958 \\ 
3.958 \\ 
${\bf \vdots }$ \\ 
3.958 \\ 
3.9501 \\ 
3.9499
\end{tabular}
& 
\begin{tabular}{c}
4.3862 \\ 
4.3857 \\ 
4.3853 \\ 
4.3852 \\ 
4.3851 \\ 
4.3851 \\ 
4.3850 \\ 
4.3850 \\ 
${\bf \vdots }$ \\ 
4.3850 \\ 
4.3812 \\ 
4.3812
\end{tabular}
& 
\begin{tabular}{c}
4.333 \\ 
4.331 \\ 
4.329 \\ 
4.327 \\ 
4.326 \\ 
4.325 \\ 
4.325 \\ 
4.324 \\ 
${\bf \vdots }$ \\ 
4.324 \\ 
4.316 \\ 
4.315
\end{tabular}
& 
\begin{tabular}{c}
4.6906 \\ 
4.6908 \\ 
4.6905 \\ 
4.6901 \\ 
4.6899 \\ 
4.6898 \\ 
4.6897 \\ 
4.6897 \\ 
${\bf \vdots }$ \\ 
4.6897 \\ 
4.6858 \\ 
4.6858
\end{tabular}
\\ \hline\hline
\end{tabular}

\item  \newpage

\item  {\bf Table 4: }Same as table 3 for $\kappa =-\,(\ell +1).$ Here we
report the Charmonium masses where the mass series and Pad\'{e} approximants 
$M_{i,\,j}$ stabilize.

\item  
\begin{tabular}{cccccc}
\hline\hline
$k,\,\ell $ & $M(N)$ & $M[i,j]$ & $k,\,\ell $ & $M(N)$ & $M[i,j]$ \\ 
\hline\hline
0,0 & M(6)=3.0333 & M[4,4]=3.0333 & 1,0 & M(7)=3.65 & M[5,5]=3.6502 \\ 
0,1 & M(5)=3.4918 & M[2,3]=3.4918 & 1,1 & M(7)=3.946 & M[4,4]=3.9462 \\ 
0,2 & M(4)=3.7787 & M[2,3]=3.7787 & 1,2 & M(7)=4.1690 & M[4,4]=4.1690 \\ 
0,3 & M(4)=4.0129 & M[2,3]=4.0129 & 2,0 & M(9)=4.08 & M[6,6]=4.0789 \\ 
0,4 & M(4)=4.2177 & M[2,3]=4.2177 & 2,1 & M(9)=4.314 & M[4,5]=4.3139 \\ 
\hline\hline
\end{tabular}

\item  \newpage

\item  {\bf Table 5: }KG results for the {\em funnel-shaped }potential $%
S(r)=br$ and $V(r)=-a/r$, with $m=1.370\,GeV$, $b=0.10429\,GeV^{2}$, $%
a=0.26, $ $E_{\hbar }$ represents the results of Kobylinsky et al [3] via $%
\hbar $-expansion (up to the third-order correction), and $E_{num}$ is the
numerical integration value reported in [3].

\item  
\begin{tabular}{cccc}
\hline\hline
$E(N)$ & $\ \ \ \ \ \ \ k=0,\,\ell =0$ & $\ \ \ \,\ \ \ \ \ \ \ k=0,\,\ell
=1 $ & $\ \ \ \ \ \ k=0,\,\ell =2$ \\ \hline\hline
\ \ 
\begin{tabular}{c}
$E(1)$ \\ 
$E(2)$ \\ 
$E(3)$ \\ 
$E(4)$ \\ 
${\bf \vdots }$ \\ 
$E(14)$ \\ 
$E_{\hbar }$ \\ 
$E_{num}$%
\end{tabular}
\ \  & \ \ \ \ \ 
\begin{tabular}{c}
1.541 \\ 
1.535 \\ 
1.534 \\ 
1.533 \\ 
${\bf \vdots }$ \\ 
1.533 \\ 
1.536 \\ 
1.533
\end{tabular}
& \ \ \ \ \ \ \ \ \ \ \ 
\begin{tabular}{c}
1.76167 \\ 
1.76064 \\ 
1.76037 \\ 
1.76033 \\ 
${\bf \vdots }$ \\ 
1.76033 \\ 
1.7604 \\ 
1.760
\end{tabular}
& \ \ \ \ \ \ 
\begin{tabular}{c}
1.90420 \\ 
1.90388 \\ 
1.90380 \\ 
1.90379 \\ 
${\bf \vdots }$ \\ 
1.90379 \\ 
1.9038 \\ 
1.904
\end{tabular}
\\ \hline\hline
\end{tabular}

\item  \newpage

\item  {\bf Table 6: }PSLET results for $\check{E}$ of equation (44) along
with the numerically predicted ones by Jena and Tripati [6] and the $1/N$%
-expansion method $E_{1/N}$ by Roy and Roychoudhury [6]. $N$ in $\check{E}(N)
$ denotes the number of corrections added to the leading term where the
series stabilizes.

\item  
\begin{tabular}{cccc}
\hline\hline
$k,\,\ell $ & $\check{E}(N)$ & $E_{num}$ & $\ \ E_{1/N}$ \\ \hline\hline
\ \ 
\begin{tabular}{c}
0, 0 \\ 
1, 0 \\ 
2, 0 \\ 
0, 1 \\ 
1, 1 \\ 
0, 2
\end{tabular}
\ \  & \ \ \ \ \ \ 
\begin{tabular}{c}
$\check{E}(2)=$1.2358 \\ 
$\check{E}(7)=$1.3347 \\ 
$\check{E}(4)=$1.3922 \\ 
$\check{E}(1)=$1.3072 \\ 
$\check{E}(4)=$1.3731 \\ 
$\check{E}(1)=$1.3540
\end{tabular}
\ \ \ \  & \ \ \ \ \ \ \ 
\begin{tabular}{c}
1.2364 \\ 
1.3347 \\ 
1.3923 \\ 
1.3071 \\ 
1.3731 \\ 
1.3544
\end{tabular}
\ \ \ \ \ \  & \ \ \ \ \ \ 
\begin{tabular}{c}
1.240 \\ 
1.340 \\ 
1.398 \\ 
1.309 \\ 
1.411 \\ 
1.358
\end{tabular}
\ \ \ \  \\ \hline\hline
\end{tabular}
\end{description}

\end{document}